\documentclass[11pt,a4paper]{article}
\pdfoutput=1

\usepackage{jheppub}
\usepackage{latexsym}
\usepackage{revsymb}
\usepackage{multirow}
\usepackage{color}
\usepackage[usenames,dvipsnames,svgnames,table]{xcolor}

\usepackage{graphicx}
\usepackage{epsfig}  
\usepackage{epsf}    
\usepackage{dcolumn}
\usepackage{bm}
\usepackage{dcolumn}
\usepackage{textcomp}
\usepackage{float}
\usepackage{hypcap}
\usepackage[]{hyperref}
\usepackage{adjustbox}
\usepackage{multirow}

\usepackage{subfigure}
\usepackage{alphalph}
\usepackage{morefloats}
\usepackage{textcomp}

\newcommand{\be}{\begin{equation}}
\newcommand{\ee}{\end{equation}}
\newcommand{\ba}{\begin{eqnarray}}
\newcommand{\ea}{\end{eqnarray}}

\newcommand{\ms}{\Delta m^2_{21}}
\newcommand{\ma}{\Delta m^2_{31}}

\newcommand{\dcp}{\delta_{\mathrm{CP}}}

\newcommand{\nova}{NO$\nu$A}
\newcommand{\mudar}{$\mu$-DAR}

\newcommand{\chisq}{\chi^2}
\newcommand{\hybrid}{T2HK($\nu$)+\mudar($\bar{\nu}$)+T2K+\nova}

\preprint{IP/BBSR/2017-6, CTPU-17-08}

\title{A hybrid setup for fundamental unknowns in neutrino oscillations using T2HK ($\nu$) and $\mu$-DAR ($\bar{\nu}$)} 

\author[a,b]{Sanjib Kumar Agarwalla,}
\author[c]{Monojit Ghosh,}
\author[d,e]{Sushant K. Raut}

\affiliation[a]{Institute of Physics, Sachivalaya Marg, Sainik School Post, Bhubaneswar 751005, India}
\affiliation[b]{Homi Bhabha National Institute, Training School Complex, Anushakti Nagar, Mumbai 400085, India}
\affiliation[c]{Department of Physics, Tokyo Metropolitan University, Hachioji, Tokyo 192-0397, Japan}
\affiliation[d]{Department of Theoretical Physics, School of Engineering Sciences, \\
KTH Royal Institute of Technology, AlbaNova University Center, 106 91 Stockholm, Sweden}
\affiliation[e]{Center for Theoretical Physics of the Universe, Institute for Basic Science (IBS), 
   Daejeon, 34051, Korea}

\emailAdd{sanjib@iopb.res.in}
\emailAdd{mghosh@phys.se.tmu.ac.jp}
\emailAdd{sushant@ibs.re.kr}

\abstract{
Neutrino mass hierarchy, CP-violation, and octant of $\theta_{23}$ are the fundamental 
unknowns in neutrino oscillations. In order to address all these three unknowns, we study
the physics reach of a setup, where we replace the antineutrino run of T2HK with antineutrinos
from muon decay at rest ($\mu$-DAR). This approach has the advantages of having higher statistics 
in both neutrino and antineutrino modes, and lower beam-on backgrounds for antineutrino run 
with reduced systematics. We find that a hybrid setup consisting of T2HK ($\nu$) and $\mu$-DAR ($\bar\nu$)
in conjunction with full exposure from T2K and NO$\nu$A can resolve the issue of mass hierarchy
at greater than 3$\sigma$ C.L. irrespective of the choices of hierarchy, $\delta_{\mathrm{CP}}$, and
$\theta_{23}$. This hybrid setup can also establish the CP-violation at 5$\sigma$ C.L. 
for $\sim$ 55\% choices of $\delta_{\mathrm{CP}}$, whereas the same for conventional 
T2HK ($\nu + \bar\nu$) setup along with T2K and NO$\nu$A is around 30\%. As far as the 
octant of $\theta_{23}$ is concerned, this hybrid setup can exclude the wrong octant at 5$\sigma$
C.L. if $\theta_{23}$ is at least $3^{\circ}$ away from maximal mixing for any $\delta_{\mathrm{CP}}$.
}

\arxivnumber{1704.06116}
\begin{document}
\maketitle

\section{Introduction and motivation}
\label{introduction}

The remarkable discovery of the Higgs boson at the Large Hadron Collider~\cite{Aad:2012tfa,Chatrchyan:2012xdj} 
marked the confirmation of the last missing piece of the Standard Model (SM) of particle physics. 
The SM explains most of the observed fundamental phenomena in particle physics with unprecedented precision. 
However, there are strong reasons to suspect that this model is not a complete description of Nature. 
For instance, in the simplest form of the SM, neutrinos are massless fermions. On the other hand, the discovery 
of neutrino flavor oscillations by the Super-Kamiokande, SNO, and KamLAND experiments~\cite{Fukuda:1998mi,Ahmad:2002jz,Eguchi:2002dm}
implies that neutrinos have mass and they mix with each other, providing an evidence for physics beyond the SM.
Several different models extending the SM have been suggested in the literature to explain the non-zero neutrino
mass and mixing (for recent reviews, see~\cite{King:2017guk,King:2015aea,Gouvea:2016shl,Altarelli:2014dca}).
High-precision measurements of the neutrino oscillation parameters can have a major impact on these models and 
can exclude a large subset of the parameter space of these models, providing crucial guidance towards our 
understanding of the physics of neutrino masses and 
mixing~\cite{King:2017guk,Gehrlein:2016wlc,Girardi:2015rwa,Ballett:2014dua,Girardi:2014faa,Albright:2006cw}.

In the three-flavor scheme, neutrino oscillation probabilities depend on six fundamental parameters. 
These are the mixing angles $\theta_{12}$, $\theta_{13}$, $\theta_{23}$, the CP-violating phase 
$\dcp$, and the mass-squared differences $\ms$ and $\ma$ ($\Delta m^2_{ij}=m_i^2-m_j^2$). 
According to the recent global fit of world neutrino data~\cite{Capozzi:2017ipn}, the solar parameters 
$\sin^2\theta_{12}$ and $\ms$ are known with a relative 1$\sigma$ precision\footnote{Here, 1$\sigma$
accuracy is defined as 1/6 of the $\pm$\,3$\sigma$ range.} of 5.8\% and 2.3\% respectively. 
The reactor angle $\sin^2\theta_{13}$ has been measured very accurately with a precision of 4\%. 
As far as the 2-3 mixing angle is concerned, the 3$\sigma$ allowed range of $\sin^2\theta_{23}$ 
is $0.381 - 0.626$, which suggests that $\theta_{23}$ can be maximal or 
non-maximal\footnote{In case of non-maximal $\theta_{23}$, it can have two solutions: one $< 45^{\circ}$,
termed as lower octant (LO), and other $> 45^{\circ}$, denoted as higher octant (HO).}.
At present, the 3$\sigma$ allowed range for $|\Delta m^2_{31}|$ is 
$2.446 \times 10^{-3}~\mbox{eV}^2 \to 2.686 \times 10^{-5}~\mbox{eV}^2$. The atmospheric mass splitting
is allowed to be either positive (known as normal hierarchy or NH) or negative (labeled as inverted hierarchy or IH)
by the present oscillation data. According to Ref.~\cite{Capozzi:2017ipn}, the current oscillation data
show an overall preference for NH with respect to IH at the level of $\sim 2\sigma$, which mainly stems
from the Super-Kamiokande atmospheric data. The complementarity among the various data sets provided
by the accelerator and reactor experiments has already enabled us to probe the parameter space for the 
$\dcp$ phase~\cite{Capozzi:2017ipn,Esteban:2016qun}. Currently, we have a hint in favor of $\dcp$ around 
$- 90^{\circ}$ and this trend is getting strengthened day by day. Also, the values of CP phase around
$\dcp \simeq 90^{\circ}$ (in the range $\sim$ $30^{\circ}$ to $ 135^{\circ}$) are disfavored at more than 
3$\sigma$ confidence level.

If in Nature, hierarchy is normal and $\dcp$ is around $- 90^{\circ}$, then the combined data from the
presently running off-axis $\nu_e$ appearance experiments, T2K~\cite{Itow:2001ee,Abe:2011ks} and
NO$\nu$A~\cite{Ayres:2002ws,Ayres:2004js,Ayres:2007tu}, can exclude the possibility of $\dcp$
being $0^{\circ}$ and $180^{\circ}$ (CP-conserving choices) at $\sim$ $2.5\sigma$ C.L. assuming
$\theta_{23}$ to be maximal~\cite{Huber:2009cw,Prakash:2012az,Agarwalla:2012bv,Ghosh:2013yon,Machado:2013kya,Ghosh:2014dba,Abe:2014tzr,Agarwalla:2016mrc}. 
Under the same conditions, these experiments can reject the wrong hierarchy at $\sim$ $3.4\sigma$ confidence level.
These experiments can also resolve the octant of $\theta_{23}$ at $2\sigma$ C.L. provided
$\sin^2\theta_{23} \leq0.44~\textrm{or}~ \geq 0.57$~\cite{Agarwalla:2013ju,Chatterjee:2013qus}.
Hence, future facilities consisting of intense high-power beams and large smart detectors
are inevitable to resolve these fundamental unknowns at high confidence level~\cite{Feldman:2013vca}.

The proposed long-baseline neutrino experiment using the Hyper-Kamiokande detector and
a powerful neutrino beam from the J-PARC proton synchrotron (commonly known as T2HK 
experiment) is one of the major candidates in the future neutrino road-map to attain the
discovery for the above mentioned issues~\cite{Abe:2011ts,Abe:2014oxa,Coloma:2012wq,Coloma:2012ji,Blennow:2014sja,Fukasawa:2016yue,Ballett:2016daj,Liao:2016orc,Ghosh:2017ged,Raut:2017dbh}.
This superbeam facility heavily relies on the data from both $\nu_{\mu} \to \nu_e$
and $\bar\nu_{\mu} \to \bar\nu_e$ channels to measure 
CP-violation\footnote{If $\dcp$ differs from both $0^{\circ}$ and $180^{\circ}$.} (CPV). 
However, dealing with the antineutrino run is very challenging for the following reasons.
The antineutrino flux is lower compared to the neutrino flux, since the production rate
of $\pi^{-}$ is less than for $\pi^{+}$. The charged current (CC) cross-section for 
antineutrino is lower than for neutrino. These two factors cause a substantial depletion
in the event rate for antineutrinos~\cite{Agarwalla:2010nn}. Due to the larger contamination 
from wrong sign pions, the background event rates are higher in case of antineutrino run and 
the systematic uncertainties are also expected to be larger~\cite{Coloma:2012ji}. 

An elegant way of tackling this issue is to replace the antineutrino run of the superbeam
experiment with the antineutrinos from muon decay at rest 
($\mu$-DAR)~\cite{Agarwalla:2010nn,Alonso:2010fs,Evslin:2015pya}.
In a stopped pion source, the low energy protons of a few GeV energy are bombarded 
on a target to produce both $\pi^{+}$ and $\pi^{-}$. Then, with the help of a high-Z beam 
stop, these pions are brought to rest. The high-Z nuclei help to capture the parent 
$\pi^{-}$ and also the daughter $\mu^{-}$. The positively charged pions undergo the
following cascade of decays at rest to produce $\nu_{\mu}$, $\bar\nu_{\mu}$, and $\nu_e$:
\begin{eqnarray*}
\pi^+ &\rightarrow &\mu^+ +\nu_\mu \\
&& \hspace{0.1cm}\raisebox{0.5em}{$\mid$}\!\negthickspace\rightarrow\ e^+ + \nu_e +\bar\nu_\mu \,.
\end{eqnarray*}
Due to oscillation, $\bar\nu_{\mu}$ can transform into $\bar\nu_e$ and these low energy
electron antineutrinos can be efficiently detected and uniquely identified with the help of
well known inverse beta decay (IBD\footnote{At energies below 100 MeV, 
the IBD cross-section is very high and also well measured. Another important feature
of this process is that it provides a very useful delayed coincidence tag between the prompt
positron and the delayed neutrino capture.}) process in a water Cerenkov detector.

The idea of combining a high energy $\nu_{\mu}$ beam from long-baseline superbeam 
facility with a low energy $\bar\nu_{\mu}$ beam from short-baseline $\mu$-DAR setup 
to measure $\theta_{13}$, mass hierarchy, and leptonic CPV was first proposed
in Ref.~\cite{Agarwalla:2010nn}. The physics reach of similar 
setups\footnote{An entirely $\mu$-DAR based experiment was considered to
measure leptonic CPV in Ref.~\cite{Conrad:2009mh}.} to measure $\dcp$ was presented 
in Refs.~\cite{Alonso:2010fs,Evslin:2015pya}. In this work, we explore the physics reach
of a setup, where we replace the antineutrino run of T2HK, with antineutrinos
from muon decay at rest ($\mu$-DAR). This setup has the benefits of having higher
signal event rates in both neutrino and antineutrino channels, and less beam-on 
backgrounds for antineutrino signal events with reduced systematic uncertainties.
We find that a hybrid setup consisting of neutrino run from T2HK and antineutrino run 
from $\mu$-DAR in combination with full data sets provided by the T2K and NO$\nu$A 
experiments can resolve all the three fundamental unknowns, namely, the 
neutrino mass hierarchy, leptonic CPV, and octant degeneracy of $\theta_{23}$ 
at high confidence level.

This paper is organized as follows. We describe the experimental setups that we have 
considered in Section~\ref{sec:setups}. In Section~\ref{sec:probability}, we 
discuss the parameter degeneracies and show how the combination of neutrino run
from T2HK and antineutrino run from $\mu$-DAR can elevate them at the probability
and event levels. This is followed by the details of our analysis in 
Section~\ref{sec:analysis}. In Section~\ref{sec:results}, we show the results of our 
simulation and discuss the features of our findings before summarizing in 
Section~\ref{sec:conclusions}.

\section{Description of experimental setups}
\label{sec:setups}

\begin{table}[htb]
 \begin{tabular}{|p{0.14\textwidth}|p{0.1\textwidth}|p{0.1\textwidth}|p{0.2\textwidth}|p{0.32\textwidth}|}
  \hline
  Experiment & Detector mass (kt) & Baseline (km) & Total pot & Systematic errors \\
  \hline
  Conventional T2HK & 560 & 295 & $7.8 \times 10^{21}$ ($\nu$) + $7.8 \times 10^{21}$ ($\overline{\nu}$) & $\nu$: 3.3\% for app and disap, $\bar{\nu}$: 6.2\% (4.5\%) for app (disap). Errors are same for sg and bg. \\
  \hline
  T2HK with \mudar & 560 & 295 & $15.6 \times 10^{21}$ ($\nu$), No $\bar{\nu}$  & Same as above for $\nu$. \\
  \hline
  \mudar\ & 22.5(SK) + 560(HK) & 15(SK) + 23(HK) & $1.1 \times 10^{25}$ ($\bar{\nu}$) (same for both SK and HK)  & 5\% for both sg and bg (same for both SK and HK). \\
  \hline
  T2K & 22.5 & 295 & $3.9 \times 10^{21}$ ($\nu$) + $3.9 \times 10^{21}$ ($\overline{\nu}$)  & sg: 2\% (0.1\%) for app (disap), bg: 5\% (0.1\%) for app (disap). Errors are same for $\nu$ and $\bar{\nu}$.\\
  \hline
  \nova\ & 14 & 810 & $1.8 \times 10^{21}$ ($\nu$) + $1.8 \times 10^{21}$ ($\overline{\nu}$)  & sg: 5\% (2.5\%) for app (disap), bg: 10\% (10\%) for app (disap). Errors are same for $\nu$ and $\bar{\nu}$.\\
  \hline
 \end{tabular}
\caption{Summary of experimental details assumed in our simulations. Where app=appearance channel, disap=disappearance channel, $\nu$=neutrino, $\bar{\nu}$=antineutrino, sg=signal and bg=background.}
\label{tab:specs}
\end{table}

T2K is a long-baseline experiment with the source of neutrinos at
J-PARC in Tokai. The detector is a 22.5 kt water Cerenkov detector 
at Kamioka, 295 km away from the source. The total beam exposure 
is $7.8 \times 10^{21}$ pot (protons on target) which is equally shared
among neutrino and antineutrino modes. We use the same configuration 
of T2K as used in Ref.~\cite{Huber:2009cw}. We have taken an uncorrelated
2\% (0.1\%) overall normalization error on signal and 5\% (0.1\%) 
overall normalization error on background corresponding to appearance 
(disappearance) channel. The systematic errors for neutrinos and 
antineutrinos are same.

For T2HK, we use the configuration as described in Ref.~\cite{Abe:2014oxa}. 
We have considered a 560 kt water Cerenkov detector with a total beam 
strength of $15.6 \times 10^{21}$ pot either running in pure neutrino mode 
or running in equal neutrino-antineutrino mode\footnote{According to the 
latest report \cite{Abe:2016ero}, the proposed detector volume for T2HK is 374 kt 
and total beam exposure is $27 \times 10^{21}$ pot. Thus, if we compare 
the exposures in terms of (kt $\times$ pot), then we see that exposures of both these 
configurations differ only by a factor of 1.2.}. The systematic errors for T2HK is 
3.3\% for neutrino mode for both appearance and disappearance 
channels and 6.2\% (4.5\%) for appearance (disappearance) channel in antineutrino mode.
Systematic errors for signal and background are same.
In our analysis, we assume that the total runs of T2HK will be completed in twelve years 
and the \mudar\ setup will be build after the first six years of operation 
of T2HK. Thus, the runtime of the \mudar\ setup is taken to be six years. 
While \mudar\ is running, we assume that the J-PARC beam continues to run 
in neutrino mode, thus doubling the neutrino exposure. 

For DAR, we consider exactly the same configuration as described 
in Ref.~\cite{Evslin:2015pya}. A six year run of \mudar\ is considered 
which corresponds to a total integrated $1.1 \times 10^{25}$ pot. 
The \mudar\ accelerator complex is assumed to be located at a distance 
of 15 km from the Super-Kamiokande (SK) detector and 23 km from the 
Hyper-Kamiokande (HK) detector. The main beam-off backgrounds for the 
$\mu$-DAR setup are invisible muons\footnote{The CC interactions 
of atmospheric muon neutrinos give rise to muons inside the detector.
The muons which are below the Cerenkov threshold will not be visible
inside the SK and HK detectors and from their decay, we will have electrons
or positrons inside the detectors which will constitute the invisible muon 
background.} and atmospheric IBD events. We have incorporated these 
backgrounds in our analysis following Ref.~\cite{Evslin:2015pya}.
A neutrino oscillation experiment with a \mudar\ source has much lower 
systematic errors, since the flux of neutrinos is exactly known from kinematics 
and the IBD cross-section is well measured in the MeV range. Here, we assume 
the systematic errors to be 5\% for both signal and background. Energy resolution 
for this set up is taken as $\delta E/E = 40\%(60\%)/\sqrt{E/{\rm MeV}}$ 
for the SK (HK) detector.

We also consider the prospective data from NO$\nu$A experiment. NO$\nu$A uses the 
NuMI beam at Fermilab with $6 \times 10^{20}$ pot/year. We have considered 
3 years running in neutrino mode and 3 years running in antineutrino mode.
The detector is a 14 kt liquid scintillator detector located 810 km away. 
We have taken a systematic error of 5\% (2.5\%) for appearance (disappearance) 
channel corresponding to signal and 10\% for both appearance and disappearance 
channels in case of background. The systematics error for neutrinos and antineutrinos 
are same. Ref.~\cite{Agarwalla:2012bv} gives the detailed experimental specifications 
for NO$\nu$A that we have used in our simulation. The experimental specifications 
of all these setups are summarized in Table~\ref{tab:specs}.

\section{Discussion at the probability and event levels}
\label{sec:probability}

\begin{figure*}
\begin{tabular}{lr}
\includegraphics[width=0.55\textwidth]{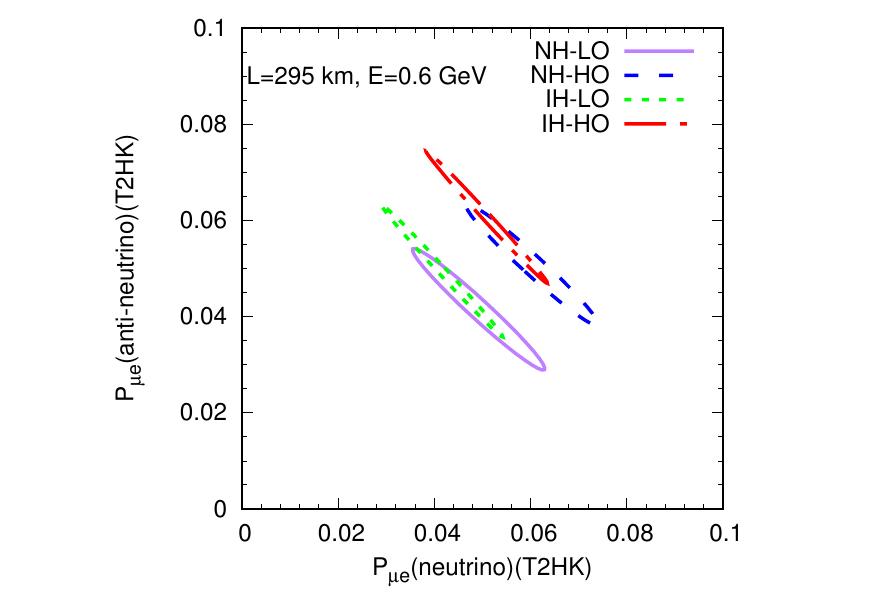}
\hspace{-0.6 in}
\includegraphics[width=0.55\textwidth]{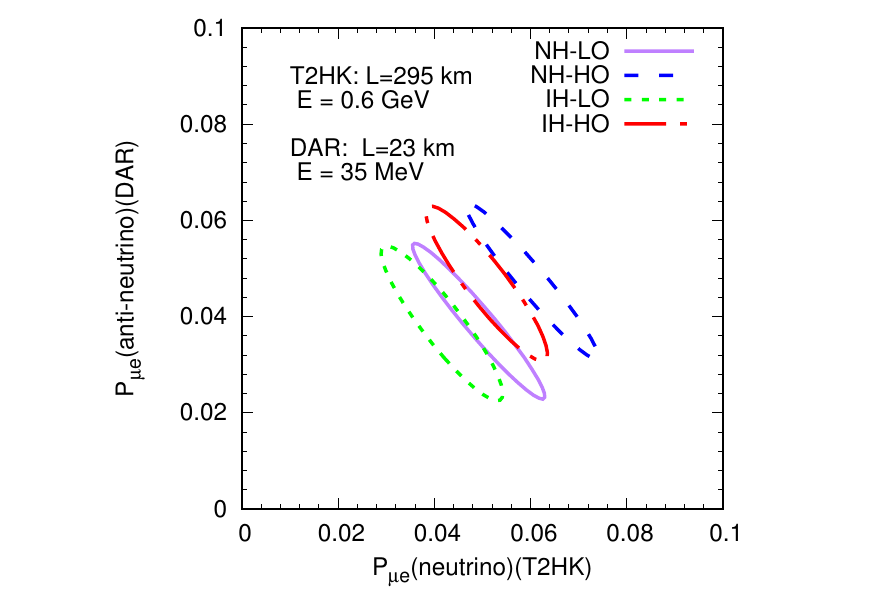} 
\hspace{-1.2 in}
\end{tabular}
\caption{$P_{\mu e}$ bi-probability plots, with neutrinos at T2HK 
on the x-axis and antineutrinos from T2HK (\mudar) 
on the y-axis in the left (right) panel.}
\label{fig0a}
\end{figure*}

\begin{figure*}
\begin{tabular}{lr}
\includegraphics[width=0.55\textwidth]{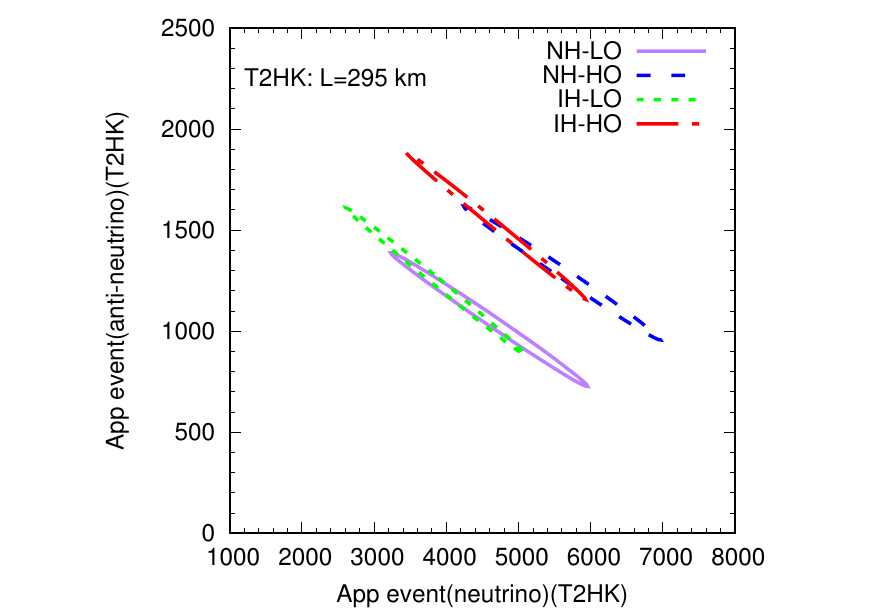}
\hspace{-0.6 in}
\includegraphics[width=0.55\textwidth]{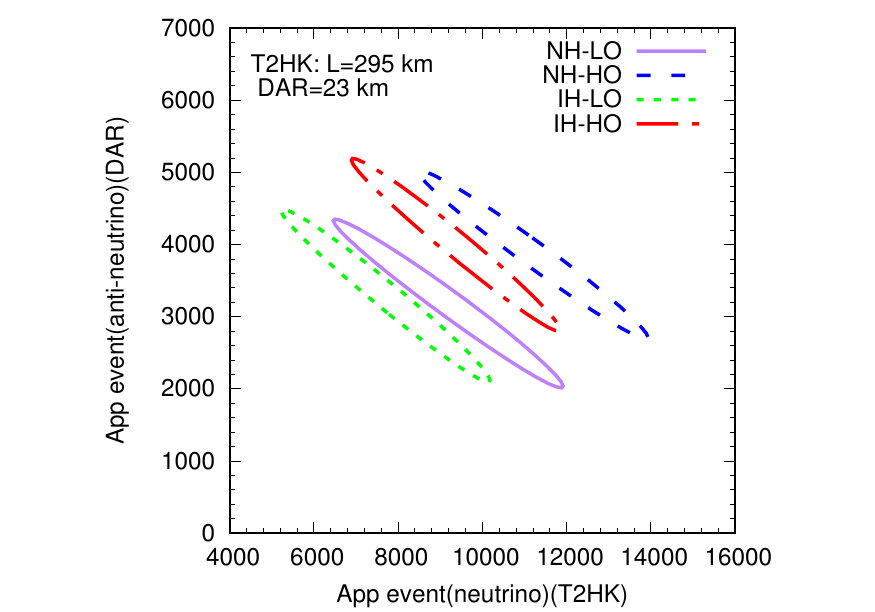} 
\hspace{-1.2 in}
\end{tabular}
\caption{$\nu_e$ appearance bi-event plots, with neutrinos at T2HK on the x-axis 
and antineutrinos from T2HK (\mudar) on the y-axis in the left (right) panel.
Note that in the hybrid setup, antineutrinos are provided by the \mudar\ source and 
the full J-PARC beam is run in the neutrino mode. That is why the neutrino events 
in the right panel are double those in the left panel.}
\label{fig0b}
\end{figure*}

The $\nu_\mu \to \nu_e$ oscillation channel is sensitive to all the unknown 
oscillation parameters. The superbeam experiments T2K, NO$\nu$A, T2HK, 
DUNE, etc. make use of this channel to determine these parameters. 
In the presence of matter and keeping terms only up to second order in
the small quantities $\alpha=\ms/\ma$ and $\sin\theta_{13}$, this oscillation
probability can be expressed 
as~\cite{Cervera:2000kp,Freund:2001ui,Akhmedov:2004ny,Agarwalla:2013tza}:
\begin{eqnarray}
\nonumber P\left(\nu_{\mu}\to \nu_{e}\right) &=& P_{\mu e} = 
\sin^2 2\theta_{13}\sin^2\theta_{23}
\frac{\sin^2[(1-\hat{A})\Delta]}{(1-\hat{A})^2} \nonumber \\
&& 
+ \, \alpha\cos\theta_{13}\sin2\theta_{12}\sin2\theta_{13}\sin2\theta_{23}\cos(\Delta+\dcp)\frac{\sin(\hat{A}\Delta)}{\hat{A}}
\frac{\sin[(1-\hat{A})\Delta]}{(1-\hat{A})} \nonumber \\
&&
+ \, \alpha^2\sin^22\theta_{12}\cos^2\theta_{13}\cos^2\theta_{23}\frac{\sin^2(\hat{A}\Delta)}{{\hat{A}}^2} \,.
\label{eq:pmue}
\end{eqnarray}
In the above equation, $\Delta=\ma L/4E$ and $\hat{A}=A/\ma$, where
$A=2\sqrt{2} G_F N_e E$ is the matter potential expressed in terms of 
the electron number density inside the Earth, $N_e$, $E$ is the neutrino energy, 
and $L$ is the baseline. The vacuum oscillation probability which is applicable 
when neutrinos travel through very little or no matter (as in the case of 
very-short-baselines) can be obtained by taking the limit $A \to 0$. 

Eq.~\ref{eq:pmue} is very useful to understand the various features of 
neutrino oscillations, especially the parameter degeneracies, and the 
resulting sensitivity of the experiments. For the purposes of this 
discussion, we neglect the term proportional to $\alpha^2$, since it 
is very small. Under a flip of the mass hierarchy, both $\alpha$ 
and $\hat{A}$ change their sign. Going from one octant to another 
affects the $\sin^2\theta_{23}$ term but not the $\sin^2 2\theta_{23}$ 
term. Finally, the effect of $\dcp$ is only felt through the term 
$\cos(\Delta+\dcp)$. As a direct consequence, we find that the 
oscillation probability for NH and $\dcp=+90^\circ$ can be matched 
by the probability for IH and $\dcp=-90^\circ$, leading to a 
degeneracy. The other two combinations are free of this hierarchy-$\dcp$ 
degeneracy. Now, going from neutrino to antineutrino 
oscillations changes the sign of both $\hat{A}$ and $\dcp$. Even if 
one looks at the antineutrino probability, the same 
degeneracy remains. Therefore, one does not expect this degeneracy 
to be lifted by the addition of antineutrino 
data~\cite{Prakash:2012az,Agarwalla:2012bv}.
Similarly, the probability for NH and LO and can be matched by 
IH and HO, leading to a degeneracy in the octant measurement. 
However for antineutrinos, the degeneracy is opposite. Thus a 
combination of neutrino and antineutrino data can lift the 
octant degeneracy~\cite{Agarwalla:2013ju}. 

In Fig.~\ref{fig0a}, we show the effect of degeneracy using the
bi-probability plots. In the left panel, for the relevant baseline 
and peak energy of T2HK, we have the neutrino (antineutrino) probability 
on the x- (y-) axis. All four combinations of the hierarchy and octant --
NH-LO, NH-HO, IH-LO and IH-HO are considered. Here, NH (IH) 
corresponds to $\Delta m^2_{31}=+(-)2.4\times 10^{-3}$ eV$^2$ and for LO (HO), 
we consider $\theta_{23} = 42^\circ (48^\circ$). As $\dcp$ varies over 
its full range, the plot gives rise to an ellipse in this plane. It is easy 
to see the hierarchy-$\dcp$ degeneracy from the overlap of NH and IH 
ellipses. As the overlap points correspond to same value 
in both neutrino and antineutrino probabilities, adding antineutrino information 
from T2HK itself cannot resolve this degeneracy. 
If we use information from one of the channels between neutrino and antineutrino, 
then the octant degeneracy arises into the picture. Since, we have informations from 
both neutrino and antineutrino channels, we can exclude the wrong octant 
irrespective of the choice of 
hierarchy ~\cite{Agarwalla:2013ju,Ghosh:2014zea,Ghosh:2015ena,Ghosh:2015tan}.
In the right panel of Fig.~\ref{fig0a}, we have replaced the antineutrino 
probability of T2HK with that of \mudar\ setup with the relevant 
baseline and energy\footnote{There are two relevant baselines 
for the \mudar\ setup -- $15$ km and $23$ km. We have chosen 
the latter baseline in the plot since it corresponds to the HK detector 
which drives the physics sensitivity by virtue of higher statistics.}.
The overlap between NH and IH ellipses for a given octant is less visible in the right panel 
as compared to the left panel. It happens due to the following reasons.
First of all, the $L/E$ informations are different in \mudar\ ($\bar{\nu}$) and T2HK ($\nu$)
setups. Secondly, the CP dependency for antineutrinos in $\mu$-DAR 
is different as compared to neutrinos in T2HK. When we combine these two setups, this fact
helps to tackle the hierarchy-$\dcp$ degeneracy. 
But, it is also important to note that the ellipses corresponding to NH-LO 
and IH-HO combinations are closer in the right panel as compared to the left panel. 
Thus, we expect that the capability of T2HK ($\nu$)+\mudar\ ($\bar{\nu}$) to resolve 
these hierarchy-octant solutions will be worse than conventional T2HK ($\nu+\bar{\nu}$).

Fig.~\ref{fig0b} is similar to Fig.~\ref{fig0a}, but at the level 
of (total) electron/positron appearance event rates. Once again 
in the left panel, we see the effect of hierarchy-$\dcp$ degeneracy 
for conventional T2HK. In the right panel, once we replace the antineutrinos 
of T2HK with antineutrinos from \mudar, we find 
that this degeneracy gets lifted. Note that the scales on the axes in 
the left and right panels are different. Due to the choice of short baseline 
and dominant IBD cross section at low energies,
the antineutrino event rates are quite high for \mudar. Moreover, since T2HK 
does not need to run in antineutrino mode, it can run in the neutrino mode 
for the entire period, doubling the neutrino event rate. Consequently, the separation 
between the ellipses is more for T2HK+\mudar\ setup.

\section{Description of numerical analysis}
\label{sec:analysis}

For our numerical analysis, we use the GLoBES package~\cite{Huber:2007ji,Huber:2004ka}
along with its associated data files~\cite{Paschos:2001np,Messier:1999kj}. 
GLoBES\footnote{GLoBES estimates the median sensitivity of the experiment 
without including statistical fluctuations. Thus, what we refer to as 
$\chi^2$ in this work corresponds to $\Delta\chi^2$ in statistical 
terminology.} calculates the simulated `true' event rates and the theoretically 
expected `test' event rates to perform a $\chisq$ analysis. 
It takes into account the statistical $\chisq$ and includes the effect 
of systematics as well as priors on the parameter values.
In our analysis we keep $\theta_{12}$ ($\sin^2\theta_{12}=0.312$), 
$\theta_{13}$ ($\sin^22\theta_{13}=0.085$), $\Delta m^2_{21} (=7.5 \times 10^{-5}$ eV$^2$) 
and $|\Delta m^2_{31}| (=0.0024$ eV$^2$) fixed
in both the true and test spectra. This is because the variation 
of these parameters in their currently allowed range does not affect 
our results much. Thus, the only relevant neutrino oscillation 
parameters in our analysis are $\theta_{23}$ and $\dcp$. 

Since the value of $\dcp$ in Nature is unknown, a good experimental 
setup must have the capability to determine the unknown parameters 
for any possible true value of $\dcp$. That is why, 
most of our results are shown as a function of true $\dcp$. 
In order to generate these results, we choose a specific 
value for true $\theta_{23}$ and $\dcp$. These two parameters 
are allowed to vary in the test range, and the relevant $\chisq$ is 
evaluated between the true and test spectra. The minimum $\chisq$ 
over the whole range of test parameters is found. This process 
is then repeated for another value of true $\dcp$ and the 
corresponding minimum $\chisq$ is evaluated. In this 
manner, we construct the plot of $\chisq$ as a function 
of true $\dcp$.

\section{Our findings}
\label{sec:results}

In this Section, we discuss the results of our numerical simulations. 
We have evaluated the sensitivity of our experimental setups 
towards determining the neutrino mass hierarchy, detecting 
CPV and determining the octant of $\theta_{23}$. The 
discussion in the following three subsections highlight our 
findings.

\subsection{Hierarchy discrimination}
\label{subsec:mh}

\begin{figure*}[t]
\begin{tabular}{lr}
\hspace{-0.8 in}
\includegraphics[width=0.52\textwidth]{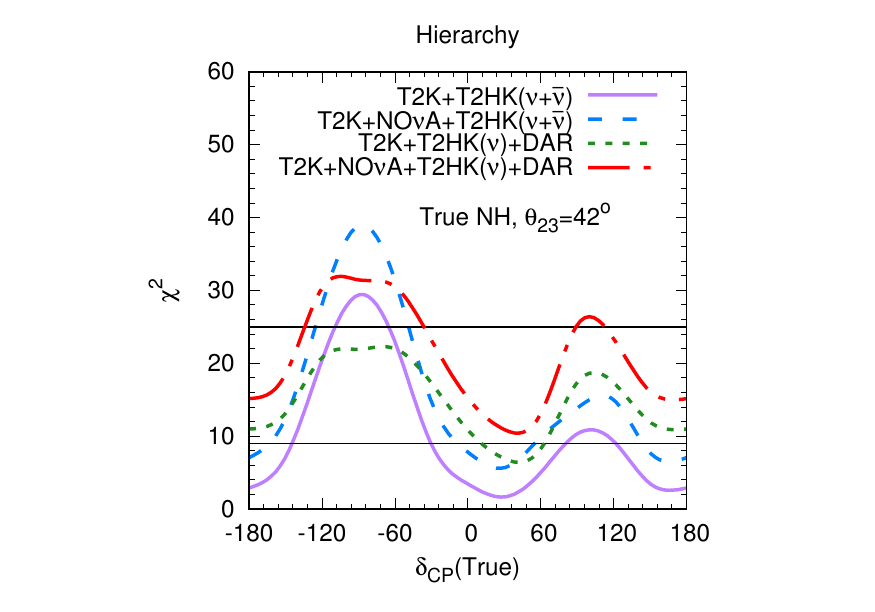}
\hspace{-1.2 in}
\includegraphics[width=0.52\textwidth]{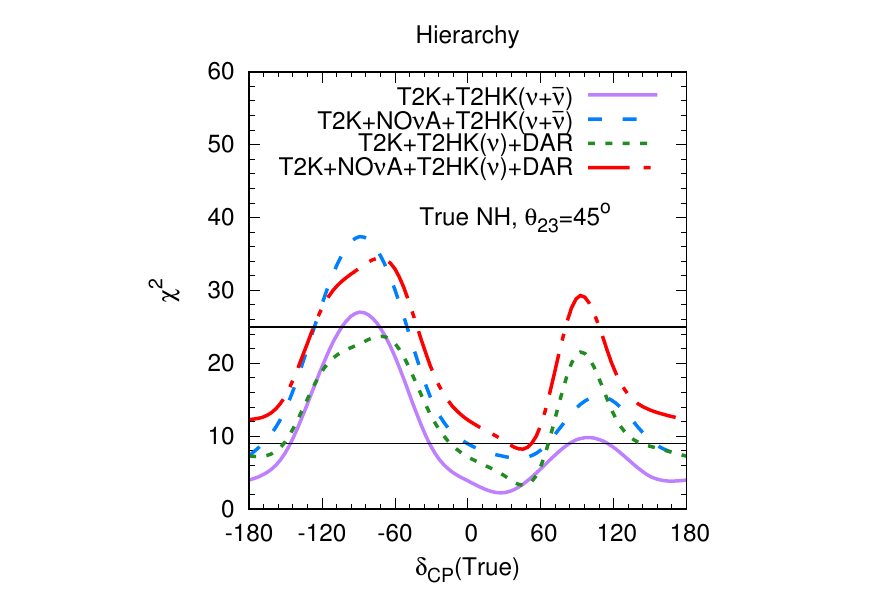} 
\hspace{-1.2 in}
\includegraphics[width=0.52\textwidth]{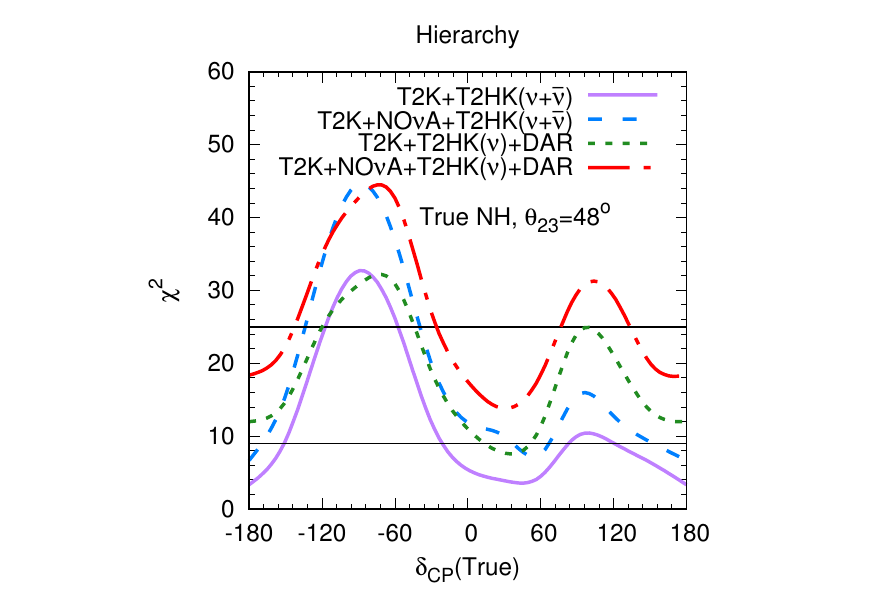} \\
\hspace{-0.8 in}
\includegraphics[width=0.52\textwidth]{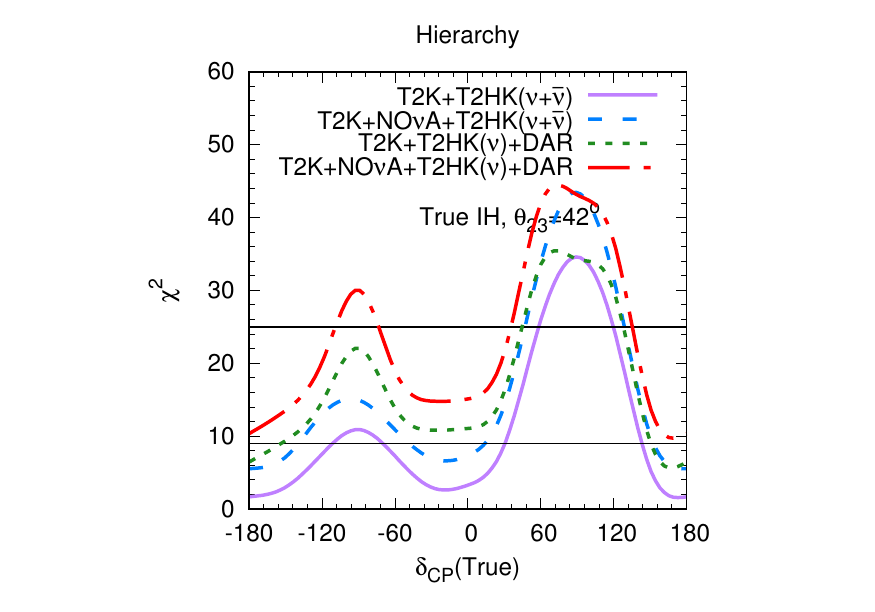}
\hspace{-1.2 in}
\includegraphics[width=0.52\textwidth]{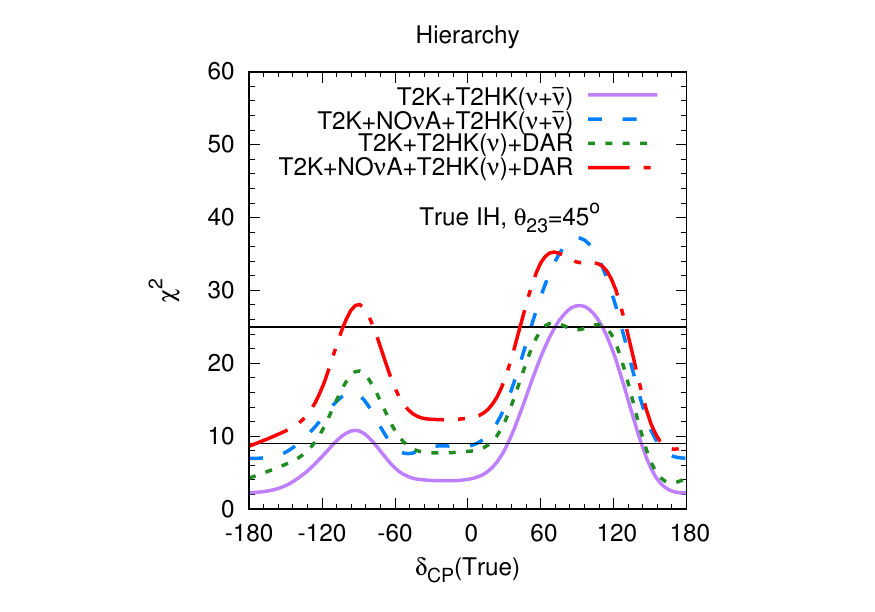}
\hspace{-1.2 in}
\includegraphics[width=0.52\textwidth]{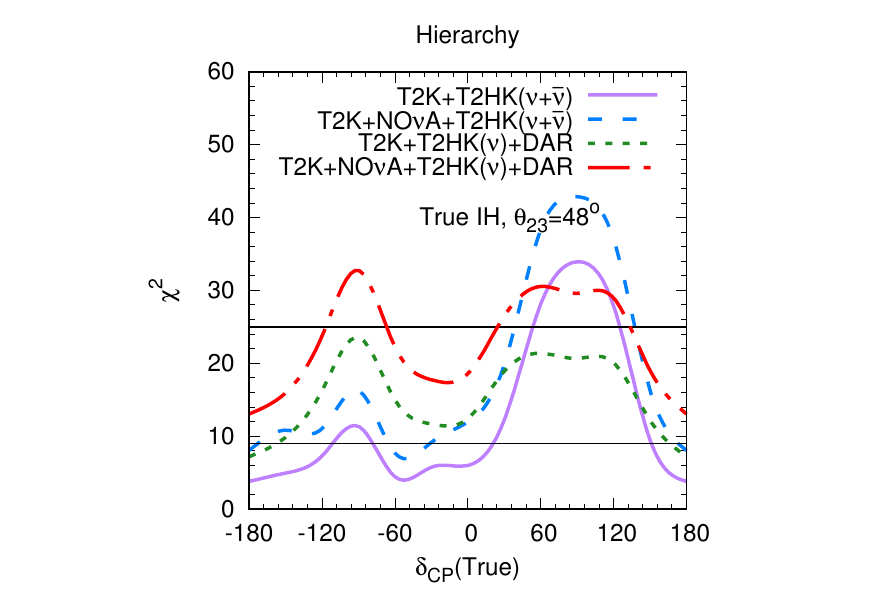}
\end{tabular}
\caption{Exclusion of the wrong hierarchy, as a function of true $\dcp$. 
The true hierarchy is NH (IH) in the upper (lower) row. The true value 
of $\theta_{23}$ is taken to be $42^\circ$/$45^\circ$/$48^\circ$ in the 
left/middle/right column. We assume that the octant is unknown, 
i.e. the test value of $\theta_{23}$ varies in both octants.}
\label{fig1}
\end{figure*}

\begin{figure*}[t]
\begin{tabular}{lr}
\includegraphics[width=0.55\textwidth]{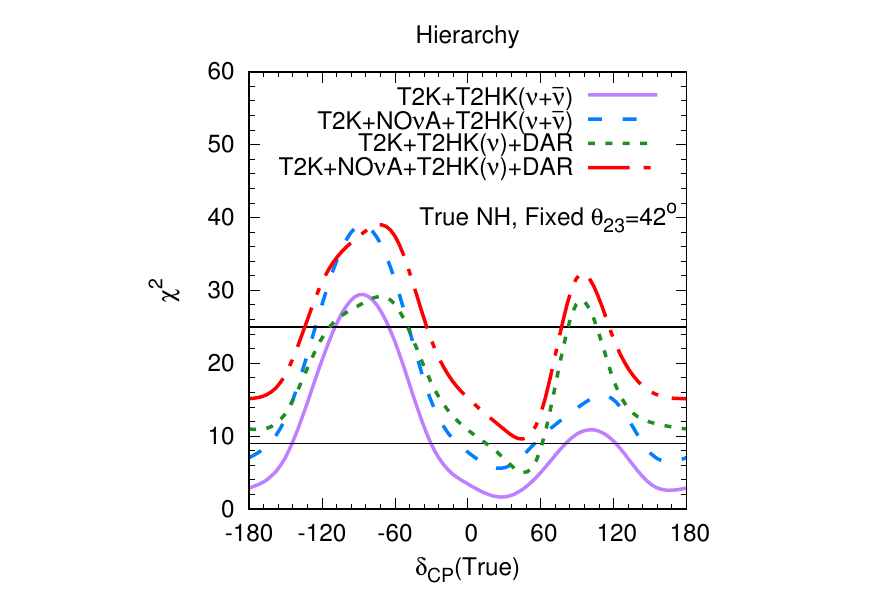}
\hspace{-1.2 in}
\includegraphics[width=0.55\textwidth]{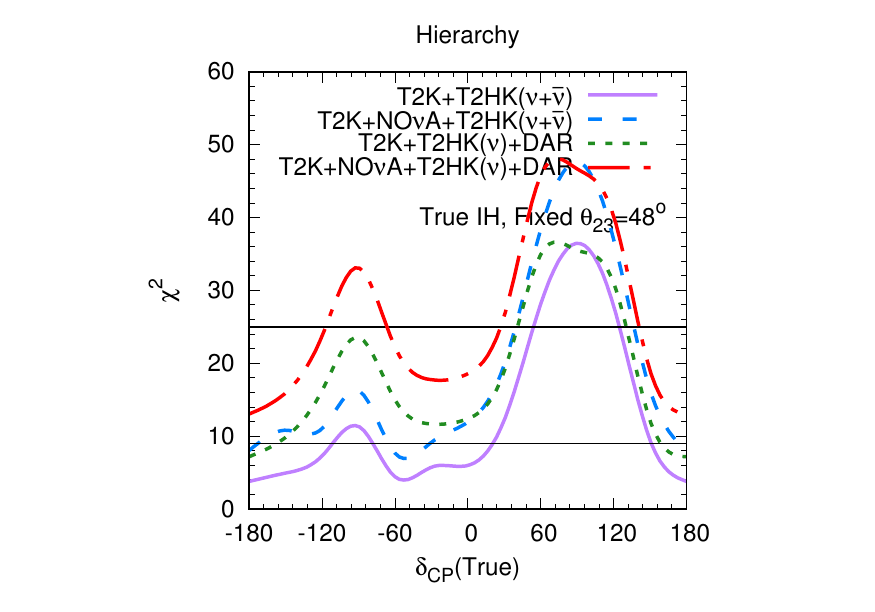} 
\end{tabular}
\caption{Exclusion of the wrong hierarchy, as a function of true $\dcp$. 
The true hierarchy and $\theta_{23}$ are specified in the panels. 
We assume that the octant is known, 
i.e. the test value of $\theta_{23}$ is varied within the same octant
as the true value.}
\label{fig2}
\end{figure*}

In this subsection, we discuss our results for determining the neutrino 
mass hierarchy. Fig.~\ref{fig1} shows the $\chisq$ for excluding the wrong 
hierarchy as a function of true $\dcp$. We assume the true hierarchy to be 
NH (IH) in the upper (lower) row. The true value of $\theta_{23}$ is taken to 
be $42^\circ$/$45^\circ$/$48^\circ$ in the left/middle/right column, 
while the test value of $\theta_{23}$ is allowed to vary in its full $3\sigma$ range. In 
each panel, we see that T2K+T2HK (violet, solid line) has good sensitivity in 
the favourable region of $\dcp$, depending on the hierarchy. In fact for 
the the combination of $\dcp=-90^\circ$ and NH which has been hinted at by 
recent T2K and \nova\ results~\cite{Abe:2017uxa,Adamson:2017gxd}, this setup can achieve well over 
$5\sigma$ exclusion of the wrong hierarchy\footnote{Here we use the 
`conventional' terminology $n\sigma$ for statistical significance, where 
$n=\sqrt{\chisq}$. For a detailed discussion on the statistical interpretation 
of oscillation experiments, we refer the reader to 
Refs.~\cite{Ciuffoli:2013rza,Blennow:2013oma,Blennow:2013kga,Elevant:2015ska}.}. Including 
data from \nova\ (blue, dashed line) improves the sensitivity further. Now, if 
we replace the antineutrino run of T2HK with \mudar\ (green, dotted line), the 
following interesting features arise. In the unfavourable range of 
$\dcp$, \mudar\ helps to lift the parameter degeneracy, which brings about an 
improvement in hierarchy exclusion. This is because the degenerate regions in 
hierarchy-$\dcp$ space for \mudar\ are different from those of T2HK, giving rise to a synergy between 
these experiments. Figure~\ref{fig1} shows that for the true 
\{hierarchy, $\dcp$, $\theta_{23}$\}-combinations of \{NH, $-90^\circ$, $42^\circ$\} 
and \{IH, $90^\circ$, $48^\circ$\}, the hierarchy sensitivity is affected 
due to the octant degeneracy. The antineutrino run of \mudar\ is not 
sufficient to lift this degeneracy and therefore the sensitivity is lowered. 
Adding information from \nova\ (red, dot-dashed line) increases the sensitivity 
further. Finally, in all the panels, the hierarchy sensitivity for the 
hybrid setup, \hybrid\ is seen to be greater than 
$3\sigma$ irrespective of the true choices of hierarchy, $\dcp$ and $\theta_{23}$. 

In order to explore the effect of octant degeneracy on our hierarchy results, we 
show in Fig.~\ref{fig2} the hierarchy exclusion results assuming the octant 
of $\theta_{23}$ is known. The two panels in this figure correspond to 
the true combinations NH-LO and IH-HO. Since the octant is assumed to be known, the test 
value of $\theta_{23}$ is only varied in the corresponding octant. 
Since the octant degeneracy is no longer relevant, we see that 
our hybrid setup \hybrid\ gives better results than T2K+T2HK, as expected.

\subsection{CP-violation discovery}
\label{subsec:cpv}

\begin{figure*}
\begin{tabular}{lr}
\hspace{-0.8 in}
\includegraphics[width=0.52\textwidth]{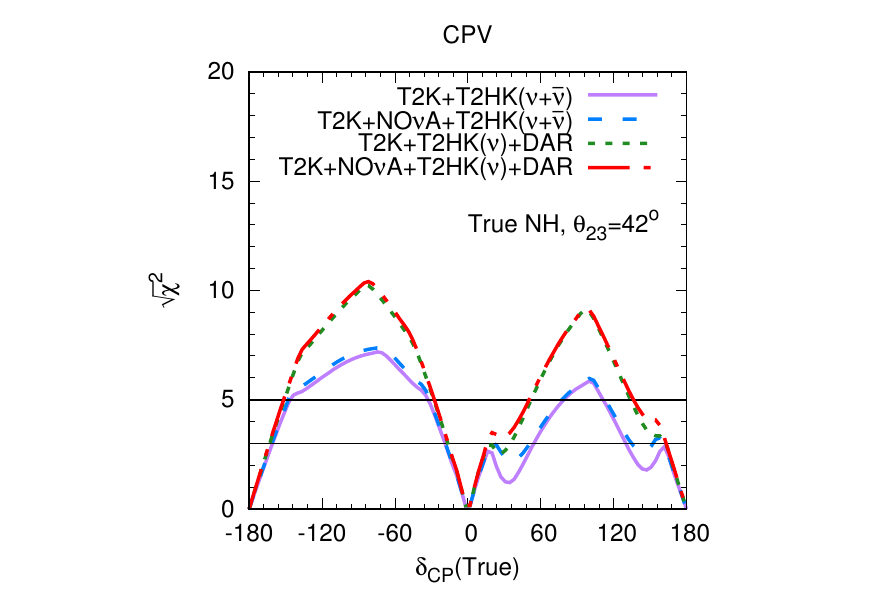}
\hspace{-1.2 in}
\includegraphics[width=0.52\textwidth]{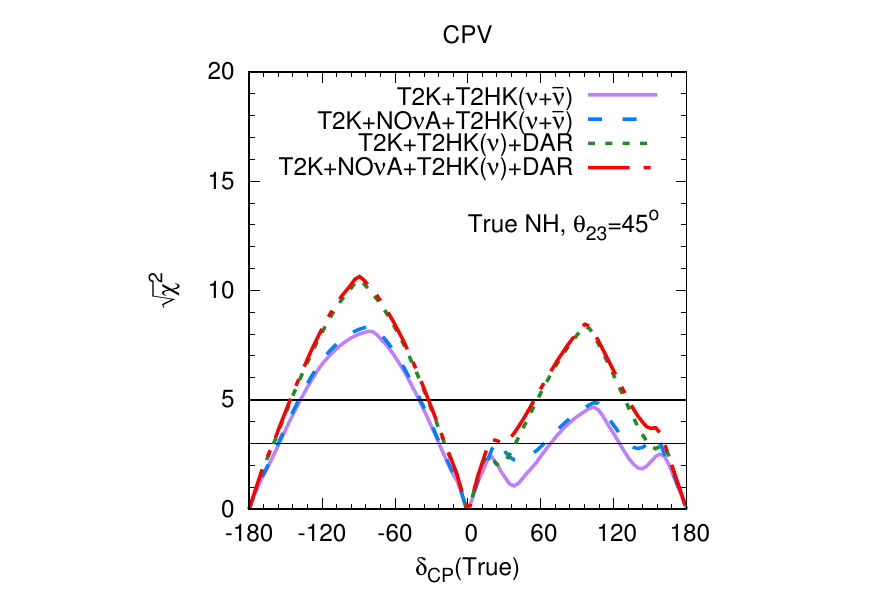} 
\hspace{-1.2 in}
\includegraphics[width=0.52\textwidth]{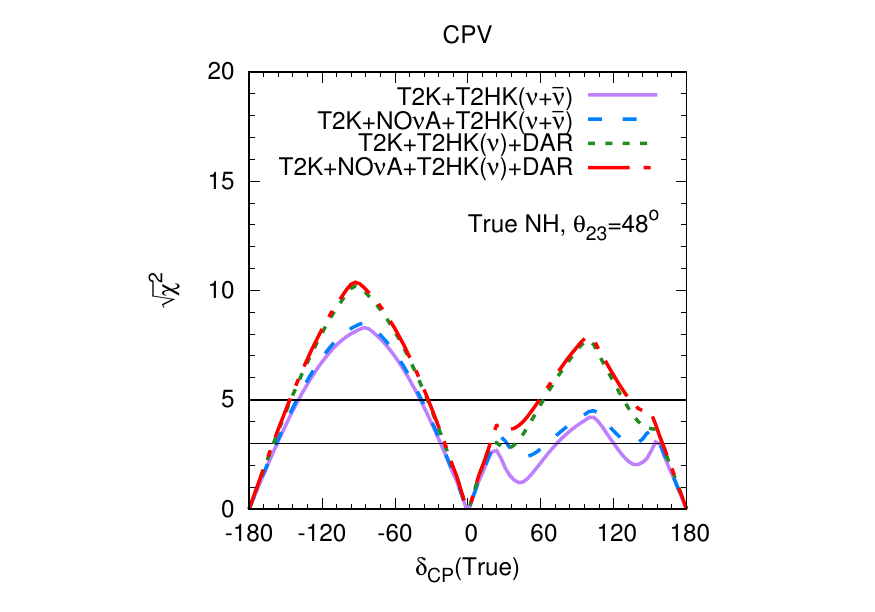} \\
\hspace{-0.8 in}
\includegraphics[width=0.52\textwidth]{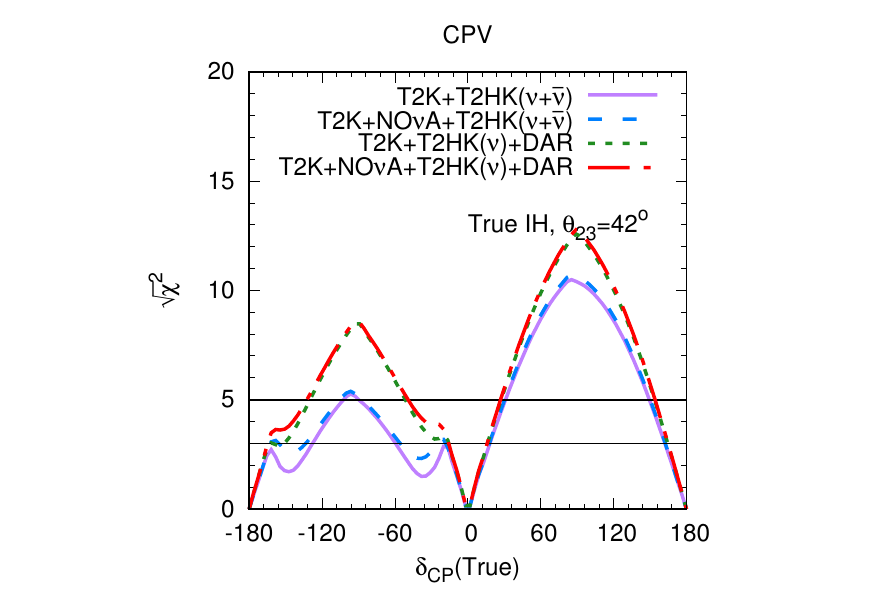}
\hspace{-1.2 in}
\includegraphics[width=0.52\textwidth]{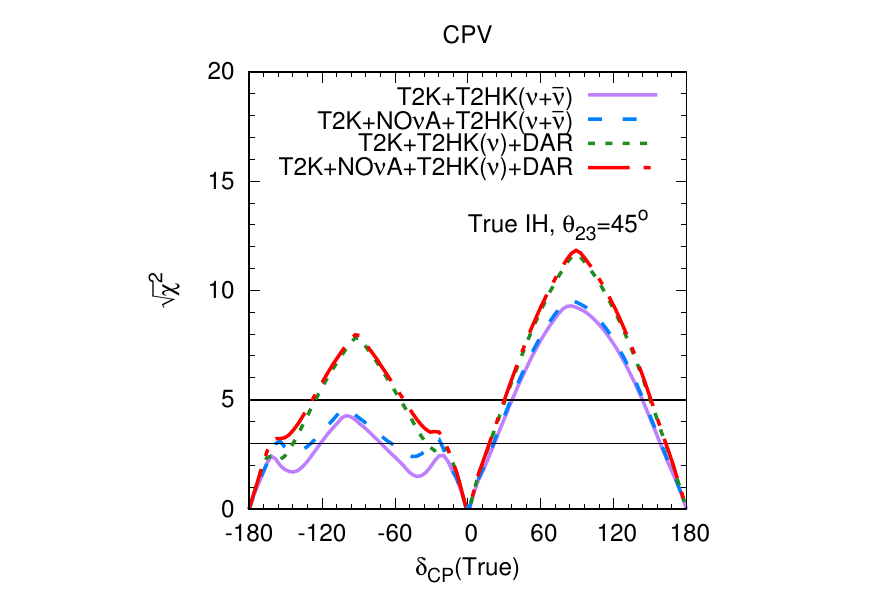}
\hspace{-1.2 in}
\includegraphics[width=0.52\textwidth]{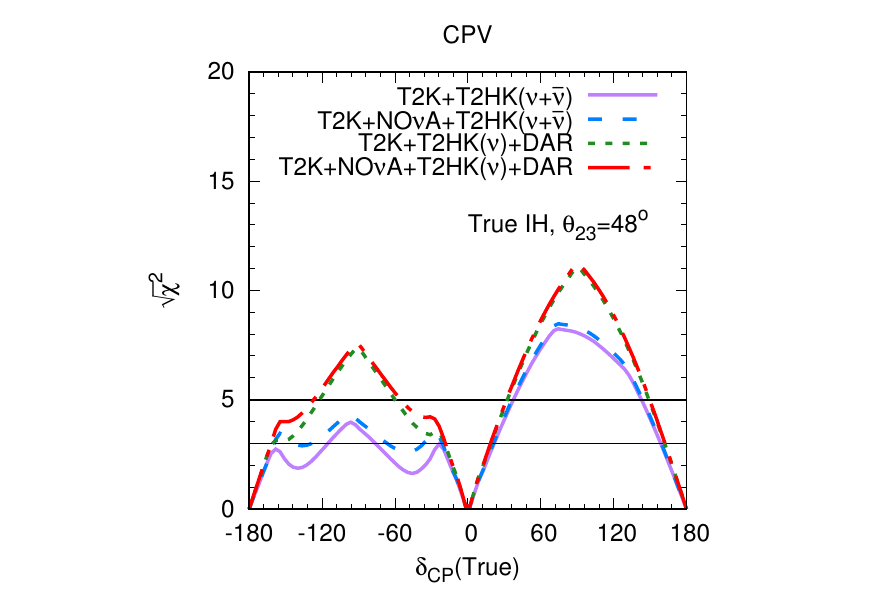}
\end{tabular}
\caption{Exclusion of the CP conserving cases $\dcp=0^\circ$ and 
$\dcp=180^\circ$, as a function of true $\dcp$. The true hierarchy is NH (IH) 
in the upper (lower) row. The true value of $\theta_{23}$ is taken to be 
$42^\circ$/$45^\circ$/$48^\circ$ in the left/middle/right column.}
\label{fig3}
\end{figure*}

Next, we discuss the CPV discovery sensitivity of the setups.
In Fig.~\ref{fig3}, we show the ability of the various setups to detect CPV 
in the neutrino sector. CP symmetry is said to be violated if the true 
value of $\dcp$ in Nature is different from $0^\circ$ and $180^\circ$. 
Thus, the sensitivity of these setups is evaluated for various values of 
true $\dcp$ in the range $[-180^\circ,180^\circ)$, while the test values of $\dcp$ are $0^\circ$ 
and $180^\circ$. We select the minimum $\Delta\chi^2$ out of these 
two test choices. The six panels of this figure correspond to the same 
true hierarchy and $\theta_{23}$ values as in Fig.~\ref{fig1}. 

Once again, we see that depending on the true hierarchy, there are 
favourable and unfavourable regions of $\dcp$ for all the setups. 
Comparing the purple (solid) and green (dashed) lines, we can clearly see a 
substantial improvement in the sensitivity when the antineutrino run of 
T2HK is replaced by antineutrinos from \mudar, and the entire runtime 
of T2HK is dedicated to the neutrino run. This holds for all 
possible combinations\footnote{Note that unlike hierarchy and octant, 
CP sensitivity does not require matter effects. 
The lack of matter effect for the \mudar\ setup along with 
its intrinsic sensitivity to $\dcp$ helps in measuring CPV.
} of the true hierarchy, octant and $\dcp$.
This improvement mainly stems from the following reasons: 
(a) the neutrino statistics for T2HK($\nu$) combined with antineutrinos 
from \mudar\ is twice compared to stand-alone T2HK($\nu+\bar{\nu}$) setup,
(b) the antineutrino event rate for \mudar\ is larger compared to 
 antineutrinos coming from the T2HK($\nu+\bar{\nu}$) setup, 
(c) the antineutrinos from \mudar\ are essentially free from any 
beam-related background, whereas the antineutrino beam in T2HK 
has substantial intrinsic beam contamination coming from 
wrong-sign mesons, and 
(d) for energies below 100 MeV, IBD provides the largest cross-section 
and comes with a useful delayed coincidence tag between the prompt 
positron and delayed neutron capture which helps in identifying 
$\bar{\nu}_e$ cleanly with high background rejection efficiency. 
Thus, antineutrinos from \mudar\ along with neutrinos from T2HK help 
to increase the fraction of $\dcp$ values for which 
CPV can be detected at a given confidence level. 
Comparing the green (dotted) and red (dot-dashed) lines 
in Fig.~\ref{fig3}, we see that adding \nova\ 
as a part of the hybrid setup doesn't help much due to its 
low statistics compared to T2HK. 

\subsection{Resolution of $\theta_{23}$ octant}
\label{subsec:octant}

\begin{figure*}
\begin{tabular}{lr}
\includegraphics[width=0.55\textwidth]{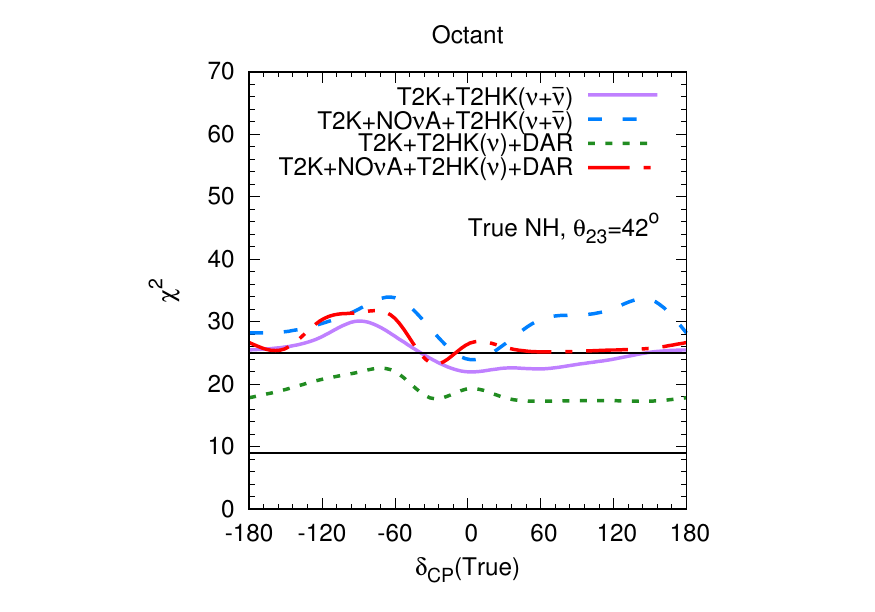}
\hspace{-1.2 in}
\includegraphics[width=0.55\textwidth]{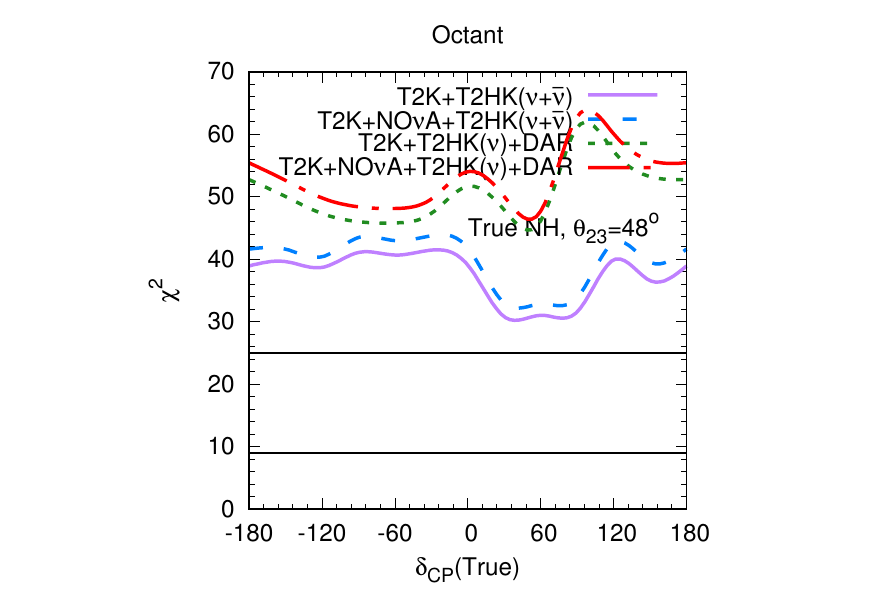} \\ 
\includegraphics[width=0.55\textwidth]{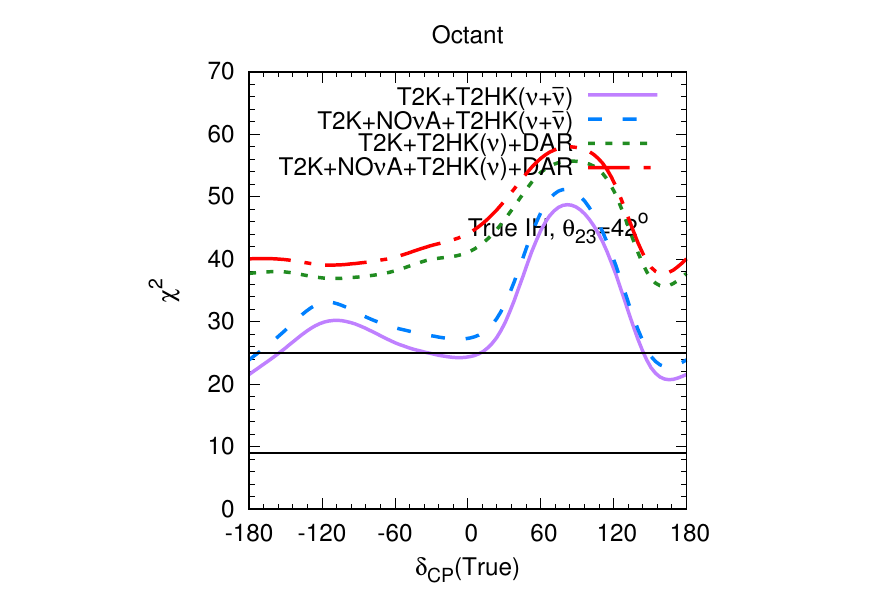} 
\hspace{-1.2 in}
\includegraphics[width=0.55\textwidth]{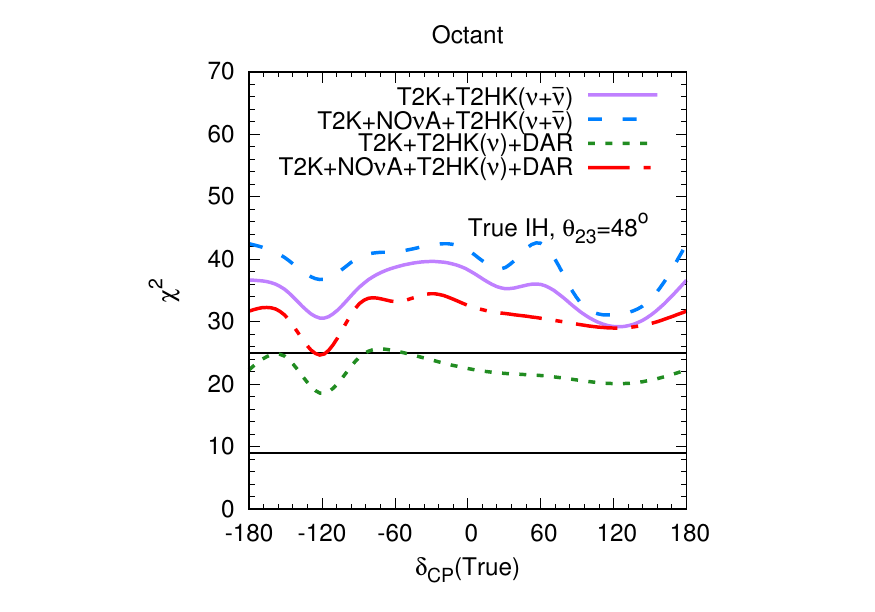}
\end{tabular}
\caption{Exclusion of the wrong octant, as a 
function of true $\dcp$. The true hierarchy is NH (IH) in the 
top (bottom) row. The true value of $\theta_{23}$ is taken 
to be $42^\circ$ ($48^\circ$) in the left (right) column. 
The true hierarchy is NH (IH) in the left (right) column. 
In all four panels, we assume that the hierarchy is unknown, 
i.e. the test hierarchy can be either NH or IH.}
\label{fig4a}
\end{figure*}

\begin{figure*}
\begin{tabular}{lr}
\includegraphics[width=0.55\textwidth]{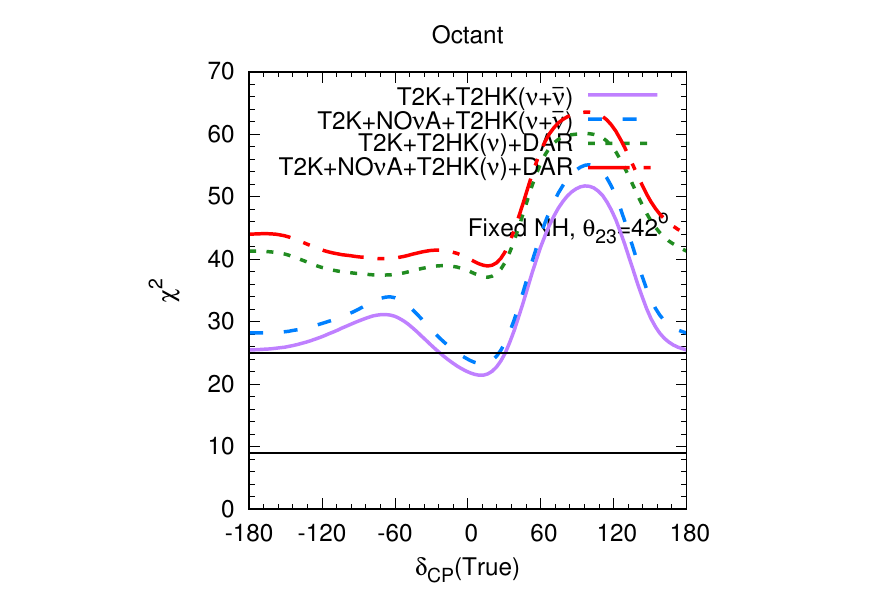}
\hspace{-1.2 in}
\includegraphics[width=0.55\textwidth]{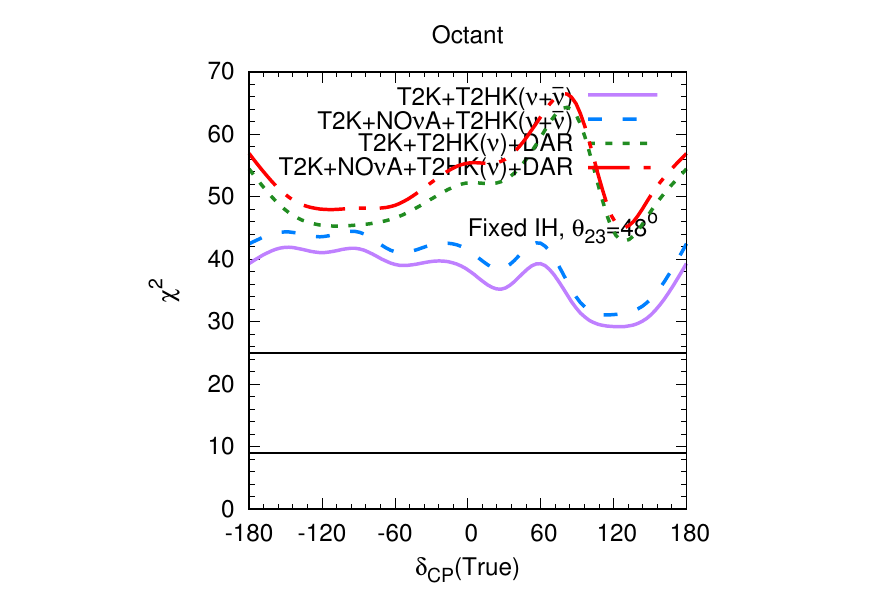} 
\end{tabular}
\caption{Exclusion of the wrong octant, as a function of true $\dcp$. 
The true hierarchy and $\theta_{23}$ are specified in the panels. 
In both panels, we assume that the hierarchy is known, 
i.e. the test and true hierarchies are the same.}
\label{fig4b}
\end{figure*}

\begin{figure*}
\begin{tabular}{lr}
\includegraphics[width=0.55\textwidth]{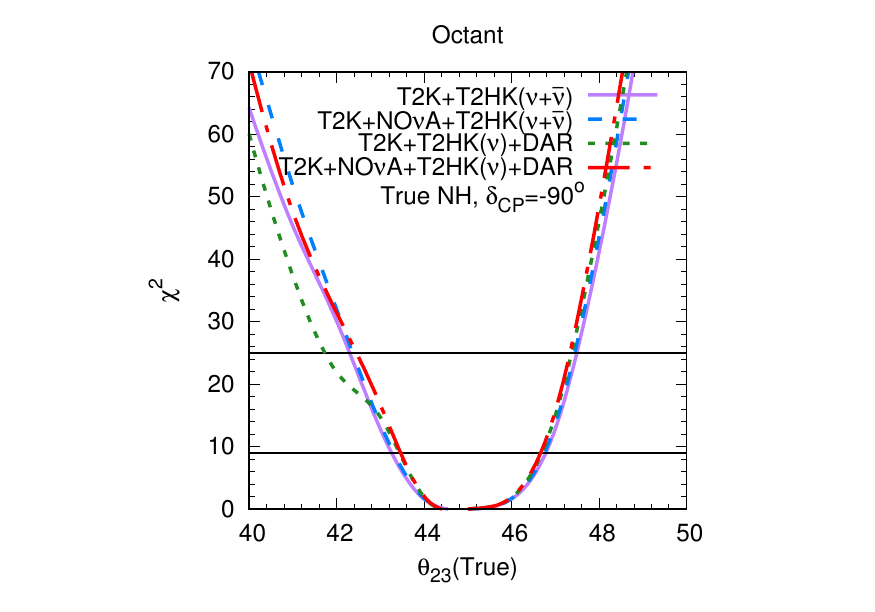}
\hspace{-1.2 in}
\includegraphics[width=0.55\textwidth]{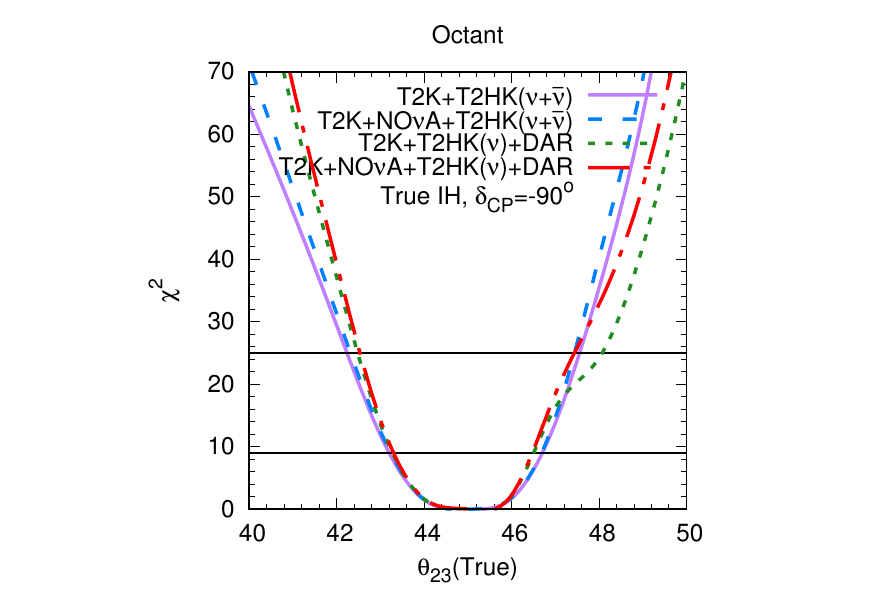}  \\ 
\includegraphics[width=0.55\textwidth]{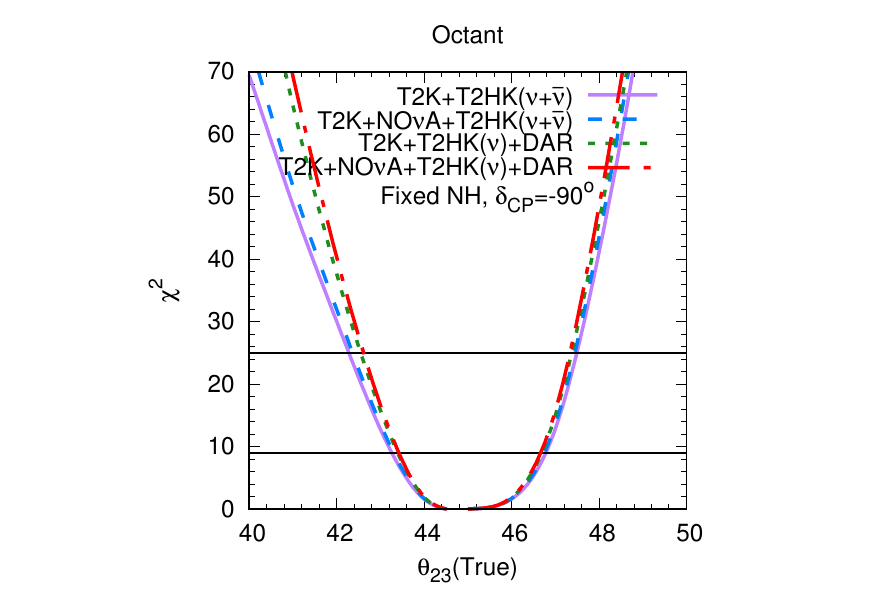}
\hspace{-1.2 in}
\includegraphics[width=0.55\textwidth]{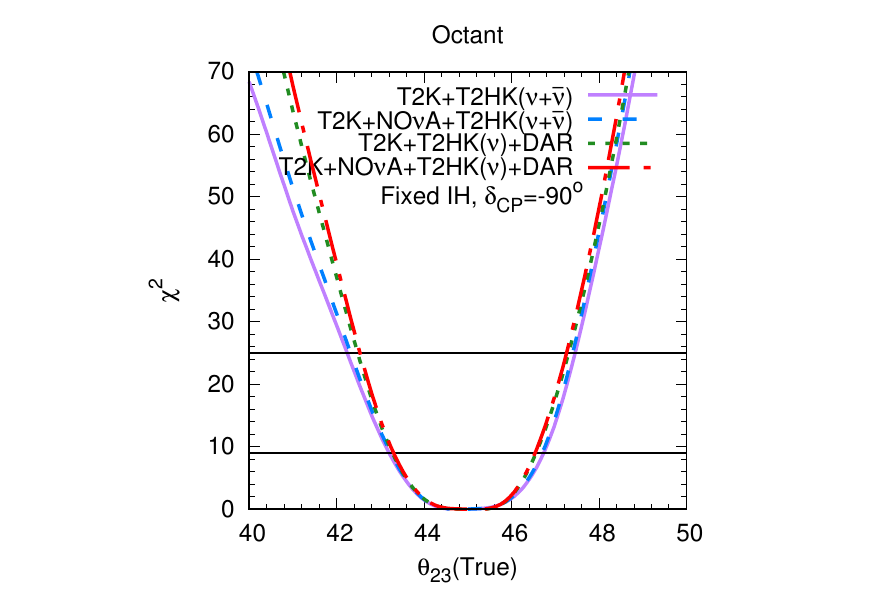}  
\end{tabular}
\caption{Exclusion of the wrong octant, as a function of true 
$\theta_{23}$, for $\dcp=-90^\circ$. The true hierarchy is NH (IH) 
in the left (right) column. We assume that the hierarchy is 
unknown (known) in the top (bottom) panels.}
\label{fig5}
\end{figure*}

Finally, we discuss the ability of our setups to determine the 
octant of $\theta_{23}$. In 
Fig.~\ref{fig4a}, we show the octant sensitivity as a function of 
the true value of $\dcp$ considering four different true 
hierarchy-$\theta_{23}$ combinations. 
Here, since we are interested in excluding the wrong octant, 
the test value of $\theta_{23}$ is varied only in  the 
opposite octant compared to the true choice. We vary test $\dcp$ 
freely throughout its entire range, and also consider both 
hierarchies in the test spectrum, i.e. we assume that the 
hierarchy is unknown in the fit.
For true NH-HO (top right panel) and IH-LO (bottom left panel), 
replacing the antineutrino run of T2HK (purple, solid line) 
by antineutrinos from \mudar\ (green, dotted line)
increases the $\chisq$ for octant determination. Addition 
of data from \nova\ helps to improve the situation further, 
as in the case of the hybrid setup (red, dot-dashed line). 

But the situation is quite different in the top left and bottom right panels. 
In these panels, the hierarchy-$\dcp$-octant degeneracy crops up in the picture, 
as discussed in Sec.~\ref{sec:probability} and antineutrinos from the 
\mudar\ source do not have enough matter effect to lift this degeneracy by 
determining the correct hierarchy. If we fix the hierarchy in the fit, this deterioration 
in the sensitivity does not occur, as demonstrated in Fig~\ref{fig4b} 
for the unfavourable hierarchy-octant combinations. Once again, 
the hybrid setup (red, dot-dashed line) is seen to outperform all the 
other setups if the hierarchy is known. 

Now, it would be interesting to see how the octant sensitivity of these 
setups varies with the different true choices of $\theta_{23}$. 
We present these results in Fig.~\ref{fig5} assuming 
$\dcp\textrm{ true}=-90^\circ$, which is close to the best-fit value as
hinted by present world neutrino data~\cite{Esteban:2016qun,Capozzi:2017ipn}.
If the hierarchy is unknown (known), the top (bottom) panels of 
this figure portray the range of true $\theta_{23}$ for which the 
wrong octant can be excluded at a given confidence level. 
All the four panels suggest that these setups can resolve the 
octant of $\theta_{23}$ at $5\sigma$ confidence level as long as 
true $\theta_{23}$ is at least $3^\circ$ away from maximal mixing. 
Again, the bottom panels confirm the fact that if the hierarchy is 
known, the hybrid setup provides the best octant sensitivity for all 
the true choices of true $\theta_{23}$. Note, we have explicitly 
checked that at $5\sigma$ C.L., the ranges of true $\theta_{23}$ 
for which the octant can be resolved remain almost the same for 
various possible choices of true $\dcp$.

\section{Conclusions and outlook}
\label{sec:conclusions}

We can explain all the current neutrino
data\footnote{We have obtained few anomalous results at 
very-short-baseline experiments which seem to point towards 
high $\Delta m^2 \sim$ 0.1-10 eV$^2$ oscillations, involving 
a hypothetical fourth mass eigenstate, which must be essentially 
sterile without having any coupling to W and Z 
bosons~\cite{Abazajian:2012ys,Palazzo:2013me,Gariazzo:2015rra,Giunti:2015wnd}.}
quite successfully in terms of three-flavor neutrino oscillations and can identify the 
remaining fundamental unknowns, in particular, the type of the neutrino mass hierarchy, 
the possible presence of a CP-violating phase, and octant ambiguity of 2-3 mixing angle. 
In this paper, we have shown that a combination of low energy $\bar\nu_{\mu}$ from 
muon decay at rest ($\mu$-DAR) facility and high energy $\nu_{\mu}$ from T2HK superbeam
experiment observed at the same Hyper-Kamiokande detector can address all these 
three pressing issues at high confidence level. This novel idea of replacing the antineutrino 
run of T2HK with antineutrinos from $\mu$-DAR yields higher statistics in both neutrino
and antineutrino modes, reduces the beam-on backgrounds for antineutrino events
significantly, and also curtails the systematic uncertainties. The low energy muon antineutrinos 
from short-baseline $\mu$-DAR setup oscillate into electron antineutrinos which can be efficiently 
detected and uniquely identified using the dominant and well known IBD process 
in water Cerenkov detector.

Our simulation shows that a hybrid setup consisting of T2HK ($\nu$) and $\mu$-DAR ($\bar\nu$)
in conjunction with full exposure from T2K and NO$\nu$A can settle the issue of mass hierarchy
at greater than 3$\sigma$ C.L. irrespective of the choices of hierarchy, $\delta_{\mathrm{CP}}$, 
and $\theta_{23}$. This hybrid setup provides a vastly improved discovery reach for CPV. 
Using this hybrid setup, we can confirm the CPV at 5$\sigma$ C.L. for almost 55\% choices of 
true $\delta_{\mathrm{CP}}$ assuming true NH and maximal mixing for true $\theta_{23}$, whereas 
the same for conventional T2HK ($\nu + \bar\nu$) setup along with T2K and NO$\nu$A is around 30\%. 
The octant resolution capability of this hybrid setup is also quite good. This hybrid setup can reject 
the wrong octant at 5$\sigma$ C.L. if $\theta_{23}$ is at least $3^{\circ}$ away from maximal mixing 
for any $\delta_{\mathrm{CP}}$. However, the practicality of this hybrid setup relies critically on the 
ongoing feasibility study of low-cost, high-intensity stopped pion sources. We hope that the results
presented in this paper in favor of this hybrid setup will boost these R\&D activities and play 
an important role in optimizing the configuration of the proposed T2HK experiment to have the best 
possible sensitivity towards determining {\it all} the remaining unknown neutrino oscillation parameters.

\subsubsection*{Acknowledgments}

S.K.A. is supported by the DST/INSPIRE Research Grant [IFA-PH-12],
Department of Science \& Technology, India.
The work of MG is supported by the €œGrant-in-Aid for Scientific
Research of the Ministry of Education, Science and Culture,
Japan€, under Grant No.  25105009. S.K.R. acknowledges support 
by IBS under the project code IBS-R018-D1.

\bibliographystyle{JHEP}
\bibliography{long-baseline-references}
  
\end{document}